\documentclass[aps,prl,twocolumn,floatfix,amsmath,amssymb,showpacs]{revtex4-1}

\usepackage{times}
\usepackage{graphicx}
\usepackage{dcolumn}
\usepackage{bm}
\usepackage{float}
\usepackage{mathrsfs}

\begin{document}

\title{Random field disorder and charge order driven quantum oscillations in cuprates}
\author{Antonio Russo}
\author{Sudip Chakravarty}
\affiliation{Department of Physics and Astronomy, University of
California Los Angeles, Los Angeles, California 90095-1547, USA}
\date{\today}

\begin{abstract}
In the pseudogap regime of the cuprates, charge order breaks a $\mathbb{Z}_{2}$ symmetry. Therefore, the interaction of charge order and quenched disorder due to potential scattering, can, in principle, be treated as a random field Ising model. A numerical analysis of the ground state of such a random field Ising model reveals local, glassy dynamics in both $2D$ and $3D$. The glassy dynamics are treated as a heat bath which couple to the itinerant electrons, leading to an unusual electronic non-Fermi liquid. If the dynamics are strong enough, the electron spectral function has no quasiparticle peak and the effective mass diverges at the Fermi surface, precluding quantum oscillations. In contrast to charge density, $d$-density wave order (reflecting staggered circulating currents) does not directly couple to potential disorder, allowing it to support quantum oscillations. At fourth order in Landau theory, there is a term consisting of the square of the $d$-density wave order parameter, and the square of the charge order. This coupling could induce parasitic charge order, which may be weak enough for the Fermi liquid behavior to remain uncorrupted. Here, we argue that this distinction must be made clear, as one interprets quantum oscillations in cuprates.
\end{abstract}
\maketitle

Recent experiments have observed a \emph{short-ranged} incommensurate charge density wave (ICDW) order in the underdoped regime of the cuprates \cite{Wu_2011,LeBoeuf_2012,Ghiringhelli_2012,Chang_2012,Hoffman_2002,Kohsaka_2007,Achkar_2012,Chang_2012,Tabis_2014}, providing a tempting explanation for the underlying order of the pseudogap regime. Here, we take a critical view of ICDW as the underlying order in terms of its ability to support quantum oscillations, which are generally agreed to reflect a Fermi surface reconstruction~\cite{Doiron_Leyraud_2007,LeBoeuf_2007,Jaudet_2008}, and therefore a Fermi liquid ground state, at least in the sense of continuity~\cite{Chakravarty:2008}. To date, there is no general agreement as to the precise nature of this reconstruction.

Because strict ICDW does not have a sharply defined Fermi surface~\cite{Zhang:2015}, there can be no quantum oscillations that are truly a periodic function of $1/B$ (where $B$ is the magnetic field). We show that even commensurate charge density wave (CDW) order---chosen for simplicity to be of period-2---in the presence of disorder may not be able to explain a Fermi surface reconstruction and consequently quantum oscillations. In short, ubiquitous potential disorder necessarily couples to CDW order, leading to a non-Fermi liquid electron spectral function without quasiparticles. If this is the case, the principal order could not be CDW.

\begin{figure}[tb]
\includegraphics[width=\linewidth]{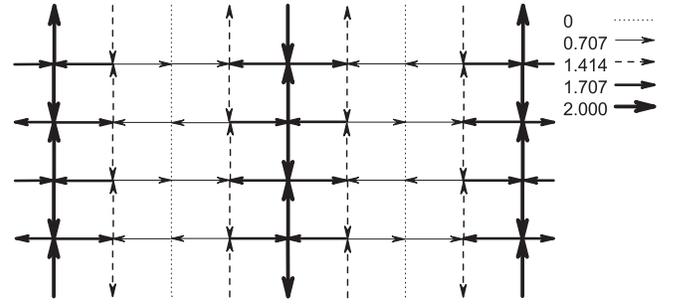}
\caption{Current pattern for period-$8$ DDW, reproduced from Ref.~\onlinecite{Eun:2012}. The wave vector $\mathbf{Q}=(\frac{3\pi}{ 4a}, \frac{\pi}{a})$, where $a$ is the lattice constant~\cite{Dimov:2008,Eun:2012}. In Landau theory, it can couple to CDW of period $2{\bf Q}$. The relative magnitudes of the currents are depicted by the thickness of the arrows in the legend. Note the antiphase domain wall structure.}
\label{fig:DDW}
\end{figure}

Another possibility for quantum oscillations is the $d$-density wave (DDW) proposed previously~\cite{Chakravarty:2008b}. This order, illustrated in Fig.~\ref{fig:DDW} in its period-$8$ version, reflects staggered, circulating currents, making it impervious to direct potential scattering. To the extent that period-8 DDW can induce period-4 CDW, the DDW can be affected by potential disorder---but only at 4th order in Landau theory. Experimentally, the situation is unclear: some neutron scattering results \cite{Mook:2002,Mook:2004} are consistent with DDW order, but nuclear magnetic resonance (NMR) measurements find no circulating currents \cite{mali:2011} (see, however, Ref.~\onlinecite{Hou:2014} for a dissenting opinion). The period-8 DDW has one electron pocket and two smaller hole pockets in the reduced Brillouin zone, thus providing an explanation of the quantum oscillations of the Hall coefficients~\cite{Eun:2012}.

For simplicity, we focus on period-2 CDW, which breaks $\mathbb{Z}_{2}$ symmetry and necessarily couples to disorder. On symmetry grounds, the effective Hamiltonian is modeled by a random field Ising model (RFIM),~\cite{ap_young-spin_glasses}
\begin{equation}
H_0 = -J\sum_{\left\langle ij\right\rangle} s_i^zs_j^z -\sum_i h_i^zs_i^z\label{classical-hamiltonian},
\end{equation}
where $J$ is the coupling between the Ising spins, and $\left\{ h_i^z\right\}$ is a set of uncorrelated, uniformly distributed (rectangular distribution) random variables with zero mean and variance $\sigma^2$. The notation $\left\langle ij\right\rangle$ denotes nearest neighbors. The model is controlled by a single dimensionless parameter, $\zeta=\frac{J}{\sigma}$.

The disorder in a RFIM drives fluctuations on many length scales, and consequently many time scales, producing glassy dynamics and a frequency-dependent susceptibility $\chi(\omega)$~\cite{Schwab_2009}. This result is recapitulated here for the $2D$ case and extended to the essential $3D$ case. Analogous to the thermally driven fluctuation of an Ising system at finite temperature, the RFIM has disorder driven fluctuations at zero temperature. A distribution $p(L)$ of domain walls of length $L$ arises from the domains in the ground state of the RFIM \cite{Schwab_2009}.

The appearance of domain walls in RFIM is identified numerically by converting the RFIM to a network flow model \cite{hartman_rieger-optimization}, and solving the ``minimum-cut'' problem, for which there are efficient algorithms \cite{Ford_1956}. Briefly, by careful choice of the parameters of the flow-network and the addition of two fictional source and sink nodes, each cut is made to correspond to a spin configuration such that the minimal cut corresponds to the RFIM ground state configuration. The probability that a domain wall of linear dimension $L$ exists in the ground state, $P_\mathrm{dw}(L)$, is determined by averaging over many disorder realizations. To help understand the meaning of this quantity, notice that, in the $1D$ case, a domain wall is just a spin flip. In the disorder-free case, creating a single spin flip costs energy $\sim 2J$, an energy that, by Jordan-Wigner transformation, can be thought of as a fermion gap. The spin flip can move throughout the system at no energy cost.

\begin{figure}
\includegraphics[width=\linewidth]{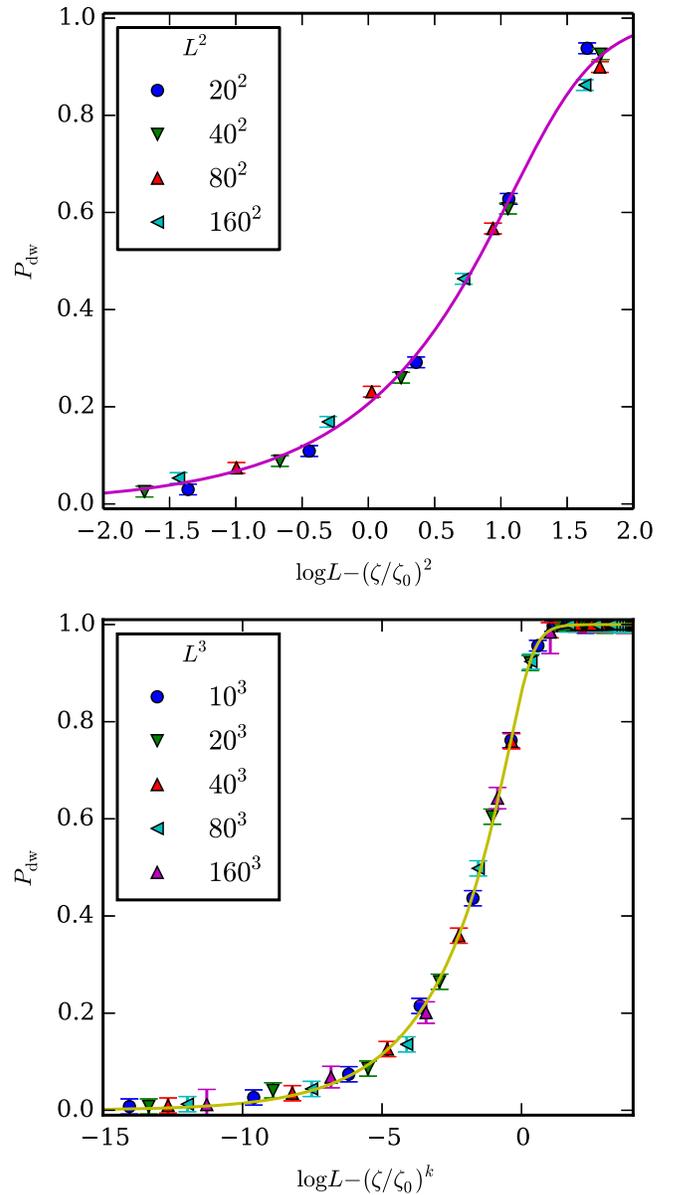}
\caption{Scaling of $P_\mathrm{dw}$ in $2D$ and $3D$.
All points are averages over $2048$ disorder realizations, except in $3D$ for $L=40$ and $L=80$, with $1024$ realizations and for $L=160$, with $512$ realizations.
\label{scaling}}
\end{figure}

In higher dimensions, the analogy is less precise, but the presence of a domain wall results in the collapse of the gap in the Ising system. Most importantly, the size of the domains is controlled by locations of these domain walls. In particular, $P_\mathrm{dw}$ is the cumulative distribution function of the ordered domains or ``clusters'' of linear dimension $L$, and therefore
\begin{equation}
p(L)= \frac{dP_\mathrm{dw}}{dL} .
\end{equation}
$P_\mathrm{dw}$ is found to lie on a universal curve~\cite{Schwab_2009} which is an asymmetric sigmoid,
\begin{equation}
P_\mathrm{dw} \approx f(x) = \frac{1}{\left(1+\exp\left[ \frac{x_0-x}{\lambda}\right] \right)^\theta}\label{cluster-distribution}
\end{equation}
where $x=\log L-\left( \zeta/\zeta_0\right)^k$ and $\zeta_0$ is numerically fit,
and sets a scale for the strength of the disorder. In $2D$, $k$ is set to $2$ as in Ref.~\onlinecite{Schwab_2009}, in agreement with the analytical result for the special case $P_\mathrm{dw}=1/2$ \cite{Binder_1983}. In $3D$, $k$ is numerically fit. The sigmoid's best fit parameters $x_0$ and $\lambda$ control its center and width, respectively, while $\theta$ controls the asymmetry. Physically, $x_0$ determines the onset of the occurrence of domain walls, and $\lambda$ how quickly the regime is dominated by the existence of at least one domain wall. The numerical results are
summarized in Fig.~\ref{scaling} and Table~\ref{fit-parameters}.

\begin{table}
\begin{tabular}{clllll}
 $D$ &	\multicolumn{1}{c}{$x_0$}&	\multicolumn{1}{c}{$\lambda$}&	\multicolumn{1}{c}{$\theta$}&
		\multicolumn{1}{c}{$\zeta_0$}&	\multicolumn{1}{c}{$k$}\\\hline
$2$&	$1.41(4)$  &	$0.28(2)$&	$0.31(2)$&	$0.75(2)$ &	\multicolumn{1}{c}{$2$}\\
$3$&	$0.2(1)$  &	$0.33(1)$&	$0.137(6)$&	$0.47(9)$ &	$5.6(1)$
\end{tabular}
\caption{Best fit parameters for $P_\mathrm{dw}$ in FIG.~\ref{scaling} and Eq.~\ref{cluster-distribution}.\label{fit-parameters}}
\end{table}

The distribution $p(L)$ is important because it controls the (necessarily local) susceptibility~\cite{sachdev_qpt,Schwab_2009}
\begin{align}
 \mathrm{Im}\, \chi(\omega) 
&\sim \int dL\, p(L) \delta\left( \omega-\omega_0 e^{-cL^\alpha}\right)
\end{align}
This phenomenological argument for the susceptibility captures the essential glassy characteristics resulting from $p(L)$. In principle, the attempt frequency $\omega_0$, the fractal dimension $\alpha$, and the length scale $c$ are microscopic parameters, which are left undetermined. Notice that the fractal dimension $\alpha\leq D$, where $D$ is the ambient spatial dimension. For $2D$ and $3D$, the integral simplifies in the small $\omega$ limit to
\begin{align}
 \mathrm{Im}\, \chi(\omega) 
\to \chi_0 \frac{\omega_0}{\omega}\Omega^\psi,\label{susceptibility-form}
\end{align}
We have put $\Omega=1/\left(\log \frac{\omega_0}{\omega}\right)$ for clarity and compactness; $\Omega(\omega)$ is strictly increasing for $0<\omega<\omega_0$, and vanishes as $\omega\to0$. The exponent
\begin{equation}
\psi=1+1/\left(\lambda\alpha\right)>1\label{exponent-psi}
\end{equation}
depends only on the fractal dimension of the domains $\alpha$ and on their distribution of sizes via the parameter $\lambda$. Moreover, in both $2D$ and $3D$, the numerical value of $\lambda$ was found to lead to $\psi>2$ (see Table~\ref{fit-parameters}).

\begin{figure}
\includegraphics[width=\linewidth]{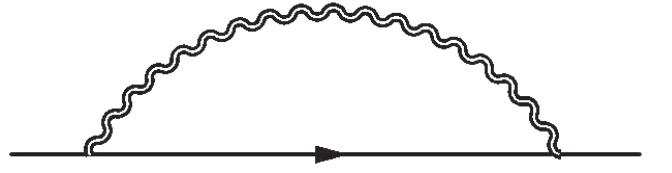}
\caption{Leading order (one-loop) self energy graph. The fermion couples to the bath of RFIM fluctuations.
\label{SEG}}
\end{figure}

We now focus on the interaction of the itinerant electrons with the emergent glassy CDW order, assumed to enter as a heat bath of fluctuations of the RFIM. The self energy $\Sigma$ of the electrons is calculated to leading order in perturbation theory (see FIG.~\ref{SEG}), assuming some coupling $\gamma$ of the RFIM fluctuations to the electrons, from the form of $\chi$ in Eq.~\ref{susceptibility-form} in a reduced graph expansion \cite{Abrahams:2014}. It is unnecessary to use the matrix formalism corresponding to the charge order, because, as we shall see, there are no quasiparticles, and hence no possible Fermi surface reconstruction. In terms of the energy of quasiparticles $\omega$,
\begin{align}
\mathrm{Im}\,\Sigma(\omega) &= -\gamma^2 \int \frac{d\omega'}{\pi} \sum_{\mathbf{q}} \mathrm{Im}\,G(\mathbf{k} - \mathbf{q},\omega-\omega')\mathrm{Im}\,\chi(\mathbf{q},\omega')\nonumber\\
&\hspace{8em}\times\left[ b(\omega')+f(\omega-\omega')\right]
\end{align}
The Fermi and Bose functions $f(\omega)$ and $b(\omega)$ restrict the $\omega'$ integration to $[0,\omega]$ in the zero-temperature limit we are considering, making the integral vanish for $\omega<0$. Because the susceptibility is local the self energy is also local, and the sum over $\bf q$ reduces to the density of states at the Fermi surface, $\nu$:
\begin{align}
\mathrm{Im}\,\Sigma(\omega)&= -\gamma^2 \nu \int_0^\omega \mathrm{Im}\,\chi(\omega')\,d\omega' \nonumber
= -\frac{\Sigma_0}{\psi-1}\Omega^{\psi-1}
\end{align}
where $\Sigma_0=\gamma^2 \nu \chi_0 \omega_0$, and $\omega>0$. From the Kramers-Kronig relations:
\begin{align}
\mathrm{Re}\,\Sigma(\omega) 
&= \frac{2}{\pi} P\int_{0}^\infty \frac{\omega'\mathrm{Im}\Sigma(\omega')}{\omega'^{2}-\omega^{2}}  \,d\omega'\nonumber\\
&= -\frac{2\Sigma_0}{\pi(\psi-1)} P\int_{0}^\Lambda \frac{\omega'}{\omega'^{2}-\omega^{2}} \Omega^{\psi-1} \,d\omega'
\end{align}
where we have introduced a cutoff $\Lambda$. Because $\Omega$ is slowly varying, we approximate it as a constant with $\omega'=\omega$:
\begin{equation}
\mathrm{Re}\,\Sigma(\omega)
\approx  \frac{2\mathrm{Im}\Sigma(\omega)}{\pi}\ln \frac{\omega_{0}}{\omega}
 = -\frac{2\Sigma_0}{\pi(\psi-1)}\Omega^{\psi-2}
\end{equation}
where we have taken the largest possible value of the cutoff, $\Lambda\to\omega_0$, and discarded the subdominant terms in the limit $\omega\to 0$. Because $\psi>2$, as $\omega\to 0$, both the real and the imaginary part of the self energy vanish.

\begin{figure}
\includegraphics[width=\linewidth]{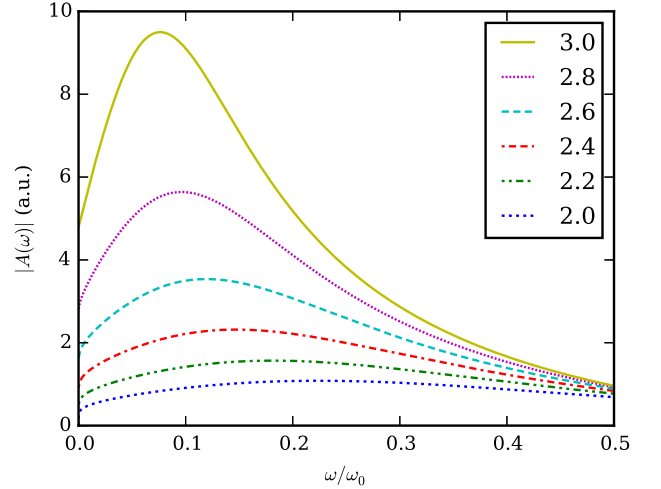}
\caption{Plot of spectral density as a function of frequency relative to the chemical potential, with wave vector $k=k_F$. Several values of the exponent $\psi$ are plotted. The relative scale of each curve is arbitrary: the unspecified prefactor $\Sigma_0$ is not included.
\label{spectral}}
\end{figure}

The spectral function $A(k=k_F,\omega)$ is plotted in FIG.~\ref{spectral} for several values of $\psi$. The emergent behavior is of an unusual non-Fermi liquid; for $k=k_F$ and in the $\omega\to0$ limit,
\begin{equation}
 A(k_F,\omega)\to \frac{\pi}{4}\frac{\psi-1}{\Sigma_0 }\Omega^{3-\psi}.\label{spectral-small-omega}
\end{equation}
Provided that $\psi<3$ or equivalently $\alpha>(2\lambda)^{-1}$ \footnote{Although we have no lower bound on $\alpha$, the condition $\alpha>(2\lambda)^{-1}$ is rather mild $\alpha\gtrsim1$ and $\alpha\gtrsim 1.5$ in $2D$ and $3D$, respectively.}, the spectral function vanishes as $\omega\to0$. The falloff is extremely slow, behaving as a fractional power of a logarithm. Furthermore, and despite the slow falloff, the quasiparticle weight always vanishes, and equivalently the effective mass diverges, as $\omega\to0$:
\begin{align}
 Z^{-1} &= 1-\mathrm{Re}\frac{\partial \Sigma}{\partial \omega}
= 1+\frac{2\Sigma_0}{\pi \omega}\frac{\psi-2}{\psi-1}\Omega^{\psi-1}
\end{align}

Cuprates are reasonably modeled as weakly coupled stacks of $2D$ layers \cite{Nie_2014}. The above work addresses isotropic coupling; we now argue that anisotropy will not materially affect the results. Consider the Hamiltonian
\begin{equation}
H_\mathrm{stacked} = -J_\parallel\sum_{\left\langle ij\right\rangle_{xy}} s_i^zs_j^z -J_z\sum_{\left\langle ij\right\rangle_{z}} s_i^zs_j^z -\sum_i h_i^z s_i^z\label{classical-hamiltonian-stacked}
\end{equation}
where $J_\parallel$ is the in-plane coupling and $J_z$ the interplane coupling. $\left\langle ij\right\rangle_z$ denotes nearest neighbors in the $z$ direction, and $\left\langle ij\right\rangle_{xy}$ the neighbors in the $xy$ plane. The random fields $h_i^z$ are as before.

Unlike $2D$, in $3D$ there is a order-disorder phase transition. In the isotropic case, i.e., $J=J_z=J_\parallel$, the zero temperature phase transition occurs at a finite $\zeta=\frac{J}{\sigma}$, found numerically to be $\zeta_c=0.446\pm0.001$, in good agreement with previous results \cite{Middleton_2002}.

\begin{figure}
\includegraphics[width=\linewidth]{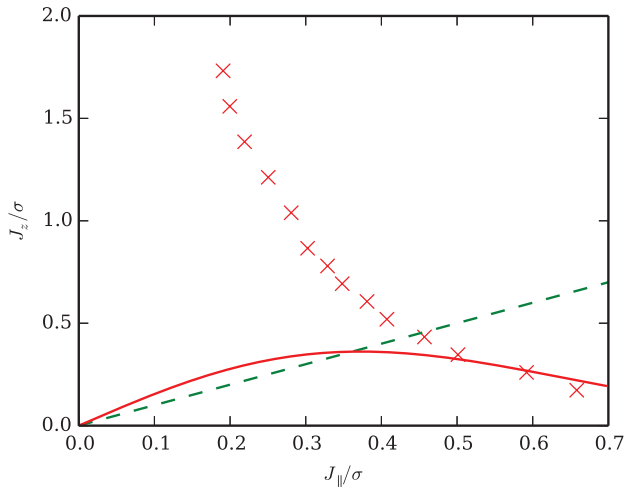}
\caption{Phase diagram for the stacked problem. The $3D$ isotropic case corresponds to the dashed diagonal line $J_z=J_\parallel$. For each value of $J_z$ considered, the numerically identified phase transition is a red $\times$. The mean field result, Eq.~\ref{meanfield-cond}, is the red line (including an approximate scale factor). The ordered phase is above the $\times$ symbols.}
\label{stacked-phase}
\end{figure}

The anisotropic case (with $J_z\neq J_\parallel$) is illustrated in Fig.~\ref{stacked-phase}. Numerically, a particular value of $J_z$ is fixed, and $J_\parallel$ is varied to identify the phase boundary in the $J_z$-$J_\parallel$ plane. A simple mean field theory result is also illustrated: as shown earlier~\cite{Schwab_2009}, in $2D$ the correlation length
\begin{equation}
\xi_{\mathrm{2D}}[J/\sigma]\sim \exp\left[ \left( \frac{J/\sigma}{\zeta_0}\right)^2 \right]
\end{equation}
with $\zeta_0\approx 0.75$. Treating the $3D$ system as a stack of coupled $2D$ planes, a mean field theory argument suggests the crossover from purely $2D$ (at weak enough $J_z$) to $3D$ occurs for
\begin{equation}
J_z\gtrsim J_\parallel/\xi_{\mathrm{2D}}^2. \label{meanfield-cond}
\end{equation}

The qualitative features are readily understood. When $J_{\parallel}\to 0$ the system decouples as a $1D$ RFIM, which cannot order (a scenario irrelevant to the cuprates). On the other hand, when $J_{z}\to 0$ the case simplifies to the $2D$ RFIM, which while it also cannot order, has an exponentially large crossover scale. It is in the latter regime that the fully $3D$ and stacked $2D$ results overlap. For weak $J_z$, the system is disordered and the total energetic contribution from the $J_z$ coupling can be made small relative to the in-plane $J_\parallel$ terms. An interpolation between the $2D$ and $3D$ cases is expected in the anisotropic case, which should always result in the supression of a quasiparticle peak.

In conclusion, random field disorder is significant even in the apparently well ordered materials of high temperature superconductors, but its effect is quite different for the two orders, CDW and DDW. Because the $\mathbb{Z}_{2}$ symmetry is broken for period-2 CDW, it is susceptible to random field disorder, destroying the Fermi surface, as we have found here by treating it as a RFIM. The disorder results in the glassy susceptibility $\mathrm{Im}\,\chi(\omega)$ of Eq.~\ref{susceptibility-form}, producing a quite unusual non-Fermi liquid. Physically, the glassy dynamics are due to the wide range of scales over which domain walls exist in the ground states of the $2D$ and $3D$ RFIM and are characterized by the parameter $\psi$, which controls susceptibility and in turn the non-Fermi liquid behavior. No Fermi-surface reconstruction can in principle occur, precluding quantum oscillations, up to some important caveats: the coupling parameter $\Sigma_{0}$ must not be too small, and the CDW correlation length cannot be too large relative to the cyclotron radius (see Ref.~\onlinecite{Tan_2015}). Truly incommensurate order in the presence of disorder is far too complex and was not addressed in the present work, but can only make things worse as far as quantum oscillations are concerned.

In contrast, DDW modulates bond currents---a Hartree-Fock calculation of DDW is given by Laughlin~\cite{Laughlin:2014}---which cannot directly couple to potential disorder, even though the order breaks translational symmetry. No non-Fermi liquid behavior is expected. Higher periodicity DDW (for example, period-8) can induce parasitic charge order that can couple to disorder. Being a higher order effect in Landau theory, this coupling may be weak. However, the observed weak CDW involves such a small motion of the atoms, it is hard to believe that it could be the cause of a large magnitude pseudogap. In any case, the short range nature of the CDW combined with RFIM disorder cannot explain quantum oscillations, at least if the resulting electronic state is a non-Fermi liquid. As a third option, if we neglect disorder and assume very long-ranged CDW, (perhaps infinitely long-ranged), Fermi surface reconstruction and quantum oscillation have been shown to be possible~\cite{Maharaj:2014,Sebastian:2014,Allais:2014,Proust:2015}. The current experiments, however, do not appear to support such assumptions.

This work was supported by a grant from the National Science Foundation, DMR-1004520. We thank E. Abrahams for discussion. We thank B. J. Ramshaw, S. E. Sebastian and D. Schwab for comments on an earlier version of the manuscript. We also thank C. M. Varma for drawing our attention to Ref.~\onlinecite{Hou:2014}.

\end{document}